\documentclass[conference]{IEEEtran}

\usepackage{cite}
\usepackage{amsmath,amssymb,amsfonts}
\usepackage{algorithmic}
\usepackage{graphicx}
\usepackage{textcomp}
\usepackage{xcolor}
\usepackage{array}
\def\BibTeX{{\rm B\kern-.05em{\sc i\kern-.025em b}\kern-.08em
    T\kern-.1667em\lower.7ex\hbox{E}\kern-.125emX}}

\begin{document}
\title{SeCoNet: A Heterosexual Contact Network Growth Model for Human Papillomavirus Disease Simulation}
\author{\IEEEauthorblockN{1\textsuperscript{st} Weiyi Wang}
\IEEEauthorblockA{\textit{Modelling and Simulation Research Group } \\ \textit{School of Computer Science} \\ \textit{Faculty of Engineering} \\
\textit{University of Sydney}\\
Sydney, Australia \\
0009-0007-9437-8862}
\and
\IEEEauthorblockN{2\textsuperscript{nd} Mahendra Piraveenan}
\IEEEauthorblockA{\textit{Modelling and Simulation Research Group} \\ \textit{School of Computer Science}\\ \textit{Faculty of Engineering} \\
\textit{University of Sydney}\\
Sydney, Australia \\
0000-0001-6550-5358}
}

\maketitle

\begin{abstract}
Human Papillomavirus infection is the most common
sexually transmitted infection, and causes serious complications
such as cervical cancer in vulnerable female populations
in regions such as East Africa. Due to the scarcity of empirical
data about sexual relationships in varying demographics, computationally
modelling the underlying sexual contact networks
is important to understand Human Papillomavirus infection
dynamics and prevention strategies. In this work we present
SeCoNet, a heterosexual contact network growth model for
Human Papillomavirus disease simulation. The growth model
consists of three mechanisms that closely imitate real-world
relationship forming and discontinuation processes in sexual
contact networks. We demonstrate that the networks grown from
this model are scale-free, as are the real world sexual contact
networks, and we demonstrate that the model can be calibrated
to fit different demographic contexts by using a range of parameters.
We also undertake disease dynamics analysis of Human
Papillomavirus infection using a compartmental epidemic model
on the grown networks. The presented SeCoNet growth model
is useful to computational epidemiologists who study sexually
transmitted infections in general and Human Papillomavirus
infection in particular.
\end{abstract}

\section{Introduction}

Human Papillomavirus (HPV) is the most common sexually transmitted viral infection; it has significant transmissibility, reaching the peak of incidences soon after the debut of sexual activity. Both men and women that are sexually active can contract the infection at least once during their lifetime. The majority infection can be cleared spontaneously; however, persistent HPV infection may progress to cancers of the anogenital area, oropharynx, neck and head, respiratory papillomatosis and anogenital warts in both men and women. Among them, cervical cancer is the most common HPV-related disease, and over 95\% of the incidences are attributed to oncogenic HPV infections. For women, persistent oncogenic HPV infection can lead to cervical intraepithelial neoplasia (CIN), which can progress to invasive cervical cancer if no responsive treatment is performed in time \cite{WorldHealthOrganization2022HumanPapillomavirus, WorldHealthOrganization2023HumanPapillomavirus}.

Modelling approaches have been utilized to simulate HPV transmission and predict the final epidemic magnitude and evaluate efficacy of prevention strategies. Most of them are developed with the infrastructures of classical models, in which the population are homogeneously mixed, and there is no demonstration of relationship among individuals. However, when it comes to STI in general or HPV in particular, infectivity is not homogeneous, and sexual contact provides the prerequisite route for infection dissemination. Complex network models can represent relationships among individuals with networks. A node in the network denotes an individual, and an edge between two nodes denote the relationship between them. Generally, it is assumed that a node's transmissibility depends on its degree that is the number of edges it has. The higher degree is, the stronger transmissibility gets. Therefore, empirical studies suggest that the underlying topology of a contact network can have a profound impact on the epidemic transmission dynamics that are transmitted through it, and the threshold of epidemic outbreaks in complex network models is different comparing to classical models \cite{Keeling2005ImplicationsNetwork}. The underlying network structure varies according to transmission routes
and infection type, and the types of networks commonly used in epidemic modelling include lattices, random networks, small-world networks and scale-free networks \cite{Harding2018ThermodynamicEfficiency}. 

The state-of-art computational HPV transmission models have mainly utilized random networks (e.g. \cite{Diez-Domingo2017RandomNetwork}) and adaptive networks (e.g. \cite{VanDeVelde2010UnderstandingDifferences, Matthijsse2015RoleAcquired}). Acedo et al. built a static lifetime heterosexual partner network in \cite{Acedo2017BuildingLifetime}, and Díez-Domingo et al. \cite{Diez-Domingo2017RandomNetwork} utilized the random network to predict HPV vaccine's long-term impact on genital warts, then Villanueva et al. \cite{Villanueva2019CalibrationAgentbased} incorporated men who have sex with men to the network. In the random network model, each node is labelled with preassigned attributes, gender, age, and expected number of lifetime partners (LSP), where LSP is assumed as a measurement for pairing. Nodes with high LSP tend to match with other nodes with high LSP. The random network avoids having nodes with extremely high degrees with constructing more hubs with fewer contacts \cite{Diez-Domingo2017RandomNetwork}. However, since the network remains static, the relationships stay active throughout the simulations without new relationship formation or relationship dissolution. Agent-based models developed by Van de Velde et al.\cite{VanDeVelde2010UnderstandingDifferences} 
and Matthijsse et al. \cite{Matthijsse2015RoleAcquired} are implemented with dynamic underlying sexual contact networks. In the former model \cite{VanDeVelde2010UnderstandingDifferences}, both casual and stable monogamous relationships are formed based on age and sexual activity level of both sides, and the dissolution depends on the female's age and sexual activity level. The latter \cite{Matthijsse2015RoleAcquired} allows concurrent relationships, and relationship formation is determined by stochastic process.

Scale-free topology is ubiquitous in social networks. Empirical studies suggest that a person's sexual attractiveness is positively correlated with people's knowledge of the given person's sexual partner amount, displaying the notion of rich-get-richer. The statistics indicate that the number of sexual partners is power law distributed \cite{DeBlasio2007PreferentialAttachment}.
Here, we propose a sexual contact network growth model that exhibits properties of scale-free network as the underlying topology of sexual contacts, then HPV transmission simulations are performed within the network. Scale-free network has not been applied in state-of-art HPV transmission models. The objective of this study is to investigate the network structure's impact on the epidemic transmission dynamics. 

The structure of the paper is organised as follows. In Section \ref{background}, we will present the background of preferential attachment models.
In Section \ref{network}, we will introduce the implementation of the sexual contact network growth model. In Section \ref{hpv}, we will introduce the HPV transmission model. The simulation results analysis will be shown in Section \ref{results}.

\section{Background} \label{background}
A network (graph) is constructed by a group of nodes (vertices) connected via edges (links). Regarding to directions of edges, networks can be divided into undirected networks and directed networks. In the former type of networks, an edge connects the two nodes on either side of the edge; while in the latter one, an edge originates from the source node and points to the target one. In this project, only undirected networks are considered.

\subsection{Degree Distribution}
A node's degree $k$ means the total number of edges that the node has. Supposing there are $N$ nodes and $M$ edges in a network, the probability of randomly selecting a node of $k$ degrees is $p_k$:
\begin{equation}\label{degree distribution}
    p_k = \frac{n_k}{N}
\end{equation}
where $n_k$ denotes the number of nodes of $k$ degrees. The degree distribution of a network demonstrates the heterogeneity of the network \cite{Piraveenan2009LocalAssortativity}.

\subsubsection{Average Degrees}
The average degree of the network can be calculated by $N$ and $M$ directly:
\begin{equation}
    \langle k \rangle = \frac{2M}{N}
\end{equation}
Alternatively, the average degree can be calculated from the degree distribution:
\begin{equation}
    \langle k \rangle = E(k) = \sum kp(k) 
\end{equation}

\subsubsection{Excess Degree Distribution}
The excess degree distribution is also called as the remaining degree distribution. As introduced above, randomly choosing a node of $k$ degree depends on $p_k$. On the other hand, the probability of arriving a node of $k$ degrees following a randomly selected edge in the network $q_k$ is different. 
\begin{equation}
    q_k = \frac{(k+1)p_{k+1}}{\sum_j jp_j}
\end{equation}
Excess degree distribution is biased by nodes of high degrees, whereas more edges are attached to a node of higher degrees rather than one of lower degrees. (\ref{degree distribution}) can be written as:
\begin{equation}
    p_k = \frac{q_{k-1}/k}{\sum_j q_{j-1}/j}
\end{equation}
Excess degree distribution does not include nodes of zero degree, since no edges are attached to them \cite{Piraveenan2009LocalAssortativity, Piraveenan2007InformationCloningScaleFree}.

\subsubsection{Joint Degree Distribution}
The joint degree distribution is defined as the probability of selecting an edge connecting a node of $i$ excess degrees to a node of $j$ excess degrees:
\begin{equation}
    e_{ij} = \frac{n_{ij}}{M}
\end{equation}
The quantity is symmetric in undirected networks; therefore, the joint degree distribution satisfies $e_{ij} = e_{ji}$, and it follows the sum rules $\sum_i e_{ij} = q_j$ and $\sum_{ij} e_{ij} = 1$ \cite{Piraveenan2007InformationCloningScaleFree, Piraveenan2009LocalAssortativity}.

\subsection{Assortativity}

Assortativity quantifies the tendency of nodes connecting with similar ones in complex networks; generally, the similarity is referred in terms of node degrees. Therefore, assortativity is defined as a correlation function of excess degree distribution and joint degree distribution of the given network:
\begin{equation}
    r = \frac{1}{\sigma^2_q}\sum_{ij}ij(e_{ij}-q_iq_j)
\end{equation}
where $\sigma^2_q$ is the variance of excess degree distribution $q_k$:
\begin{equation}
    \sigma^2_q = \sum_kk^2q_k - {[\sum_kkq_k]}^2
\end{equation}
The value of assortativity ranges from -1 to 1. $r = 1$ denotes that the network has absolute assortativity, where nodes only connect with nodes of the same degree. $r = 0$ denotes that the network has no assortativity (random linking), where nodes are randomly linked to other nodes \cite{Piraveenan2007InformationCloningScaleFree, Piraveenan2009LocalAssortativity, Lizier2011FunctionalStructural, Law2020PlacementMatters}.

\subsection{Scale-free Networks}
Scale-free topology is ubiquitous in social, biological and technical networks, exhibiting power-law degree distribution that the probability mass function (pmf) of degree $P(k)$ in the network is distributed following the functional form:
\begin{equation} \label{power law}
    P(k) = Ak^{-\gamma} 
\end{equation}
where $A$ is a coefficient, and $\gamma$ is known as scale-free exponent (also referred as the power law exponent), whose value typically ranges from 2 to 3. A flatter degree distribution slope corresponds to a lower value of $gamma$, while a higher value of $gamma$ results in a steeper slope in the degree distribution. Attachment growth models are developed to generate scale-free networks \cite{Bell2017NetworkGrowth, Law2020PlacementMatters}.

\subsubsection{Preferential Attachment Model}
Preferential attachment (PA) models have been the most prominent approach to generate scale-free networks. The term of PA was introduced by Barabási and Albert \cite{Barabasi1999EmergenceScaling}, who proposed the most well-known preferential attachment model, the Barabási-Albert (BA) model. In the model, the initial connected network consists of $m_0$ nodes. At each time step, a new node is introduced to the network and is connected to $m < m_0$ existing nodes in the network. The probability $p_i$ that a new node is attached to an existing node $i$ in the network is proportional to its degree $k$, determined as:
\begin{equation} \label{ba}
    p_i = \frac{k_i}{\sum_{j\in N}k_j}
\end{equation}
where $k_i$ is the degree of node $i$, and $N$ is the set of preexisting nodes $j$ available to a new edge in the network. Therefore, nodes with high degrees are more likely to receive new edges, exhibiting a rich-get-richer mechanism. Meanwhile, as the network expands with time, the attachment probability is also correlated to the duration that the node has existed in the network \cite{Bell2017NetworkGrowth}.

\subsubsection{Fitness Model}
Besides node degree, node fitness has been an alternative notion in attachment growth models. A node's fitness can be explained as the attractiveness of the node, which decides the rate of receiving edges from other nodes as the network evolves over time. It can be a single attribute of the node such as the degree, which corresponds to the BA model, or the attribute amalgamation of the given node that contributes to its propensity to attract new neighbors. 

A prominent example emphasizing on fitness itself and eliminating preferential attachment mechanism is the Caldarelli model \cite{Caldarelli2002ScaleFreeNetworks}. In the beginning, there are a fixed number $N$ of nodes, and each node is randomly assigned with a fitness value $x_i$ from a given probability distribution $\rho(x)$. Considering the probability of connecting two nodes $i$ and $j$ is
\begin{equation}
    f(x_i, x_j) = \frac{(x_ix_j)}{x^2_M}
\end{equation}
where $x_M$ is the maximum value of $x$ in the network. The mean degree of nodes with fitness $x$ becomes
\begin{equation}
    k(x) = \frac{Nx}{x^2_M}\int^{\infty}_0y\rho(y)dy = N\frac{\langle x \rangle x}{x^2_M}
\end{equation}
The probability distribution of $P(k)$ is 
\begin{equation}
    P(k) = \frac{x^2_M}{N\langle x \rangle}\rho(\frac{x^2_M}{N\langle x \rangle}k)
\end{equation}
As a result, both the fitness distribution and the degree distribution follow the power law \ref{power law}, which indicates that preferential attachment is not the only underlying mechanism for power law degree distribution.

Fitness can be a static attribute throughout the growth process, or a dynamic attribute that changes over time. Fitness based preferential attachment models ensures that newly added nodes with high fitness are also competent to receive new connections, which is more associate with the reality, for example, new publications receive multiple citations in a short time \cite{Bell2017NetworkGrowth}.

\subsubsection{Fitness-based Proportional Model}
Fitness-based proportional attachment models combines the two approaches mentioned above. The models incorporate 'preferential attachment' from models as the BA model with 'fitness' from models as the Caldarelli model. The integrated mechanism ensures that when the underlying fitness distribution does not display power law, the scale-free topology can still emerge \cite{Bell2017NetworkGrowth}.

A typical example of fitness-based proportional attachment models is the Bianconi-Barabási model \cite{Bianconi2001CompetitionMultiscaling}, which is generally seen as a variant of the BA model. The model also starts with a network consisting of $m_0$ connected nodes, and at each time step, a new node is introduced to the network and is connected to $m < m_0$ existing nodes in the network. Additionally, each node $i$ in the network is assigned with a fitness value $\phi_i$. The probability $p_i$ of a new node linking to an existing node $i$ in the network is decided by both degree and fitness of node $i$:
\begin{equation}
    p_i = \frac{k_i\phi_i}{\sum_{j\in N}k_j\phi_j}
\end{equation}
Thus, when two nodes $i$ and $j$ share the same degree ($k_i = k_j$), the one with higher fitness is more competent to attract a new connection. Similarly, when two nodes $i$ and $j$ have the same fitness ($\phi_i = \phi_j$), the one with higher degree is more likely to receive a new connection. Comparing to the BA model, it is possible for a relatively young node in the Bianconi-Barabási model to overtake an older one to attach to more new neighbors.

\subsection{HPV and Cervical Cancer}
HPV infects skin or muscosal cells, and there are over 100 recognised genotypes of the virus. Among them, the ones that can cause cancer of the cervix or are associated with cancers of other areas including head, neck, oropharynx the anogenital area are known as the "high-risk" types, and the rest are known as the "low-risk" types. HPV 6 and 11 are the commonly known low-risk genotypes, which can cause anogenital warts on the external genitalia with high morbidity. For high-risk genotypes, HPV 16 and 18 are responsible for approximately 70\% of all cervical cancer incidences \cite{WorldHealthOrganization2023HumanPapillomavirus}. 

Cervical cancer is the most common HPV-related disease, and it is also the fourth most frequent malignancy among women worldwide, which poses a severe threat to public health. The diagnosis of CIN and cervical cancer is attributed to persistent high-risk HPV infection. It generally takes 15 to 20 years for oncogenic HPV infection to become chronic and progress to precancerous lesions then further carcinoma with normal functioning immune system; however, if the immune system is damaged or weakened due to other sexually transmitted infections (STI) or unhealthy living habits, the whole procedure may reduce to only 5 to 10 years \cite{WorldHealthOrganization2022CervicalCancer}.

With HPV vaccination and screenings, HPV infection is preventable. All current licensed vaccines target at HPV 16 and 18, and the suggested primary recipient population is teenage girls aged 9 years or older before their first sexual intercourse, for which, after the debut of sexual activity, young women are at high risk of contracting HPV infection. Meanwhile, screening can identify early HPV infection and precancerous lesions \cite{WorldHealthOrganization2022CervicalCancer, WorldHealthOrganization2022HumanPapillomavirus}.

\section{Methodology}

Both the sexual contact network growth model and the HPV transmission model are implemented using the software Netlogo \cite{Wilensky2021NetLogo}. 

\subsection{The SeCoNet Sexual Contact Network Growth Model} \label{network}

We propose a growth model which will essentially create scale-free networks, with constraints which are specific to a sexual contact network. We focus on heterosexual contacts only in this study, thus the grown networks will be bipartite networks.  The justification for creating bipartite scale-free networks to model  heterosexual contact networks is as follows. The  number of sexual contacts is not distributed homogeneously among people. People have personal attributes, or perceived attributes, such as age, income, appearance, education, social status etc which make some people more attractive as sexual partners, resulting in a heterogeneous distribution in terms of number of sexual contacts.  The involvement of career choices such as being sex workers and perspectives on commercial sex also impact the Lifetime Sexual Partners (LSPs) a person will accumulate. Furthermore,  flirting skills  typically improve with practise, potentially increasing the success probability of pick-ups in return, which corresponds to a rich-gets-richer mechanism \cite{Merton1988MatthewEffect} which typically results in the creation of scale-free networks. It has also been demonstrated that real world sexual contact networks exhibit scale-free topology \cite{DeBlasio2007PreferentialAttachment}.  For these reasons,  we propose a sexual contact network growth model which will result in the creation of scale-free networks with characteristics particular to heterosexual contact networks.  We name this the SeCoNet (\textbf{Se}xual \textbf{Co}ntact \textbf{Net}work) Growth Model, which is described below. The model at present uses attributes typical to the Australian population, but can easily be calibrated to be applicable to other populations as well.

\subsubsection{Initialisation}

The SeCoNet  growth model generates a scale-free heterosexual network consisting of $N$ individuals and their relationships,  in order to capture the individual heterogeneity in sexual contact dynamics, where the individuals are denoted by nodes, and the relationships are represented by edges in the network. Each individual (node) $i$ is assigned a set of characteristics and attributes at initialisation: (1) age $g_i$; (2) gender (1 for females and -1 for males); (3) the estimated average duration of relationships $\delta_i$.  The individuals are assumed to be aged between 15 to 49 years old. The age distribution is based on the age distribution of Australia as described in the 2021 Australian census \cite{AustralianBureauOfStatisticsCensusOfPopulationAndHousing2021FiveYear} , which is summarised in Table \ref{age}.

\begin{table}[htbp]
\caption{Age distribution used in SeCoNet Growth Model}
\begin{center}
\begin{tabular}{|c|c|c|c|}
\hline
 Age Group & \% & Age Group & \% \\
\hline
15 to 19 & 12.3 & 35 to 39 & 15.6 \\
20 to 24 & 13.4 & 40 to 44 & 14.0 \\
25 to 29 & 15.1 & 45 to 49 & 13.8 \\
30 to 34 & 15.8 &  &   \\
\hline
\end{tabular}
\label{age}
\end{center}
\end{table}

\begin{table*}[htbp]
\caption{List of variables and parameters for the SeCoNet growth model}
\begin{center}
\begin{tabular}{|c c m{5cm} m{4.5cm}|}
\hline
Variable / Parameter & Symbol & Value & Source \\
\hline
\textbf{Variable} & $ $ & &   \\
     & $ $ & &   \\
Timestep & $t$ &  &   \\
Number of contacts & $M$ &  & \\
Age  & $g_i$ &  & Australia Bureau of Statistics \cite{AustralianBureauOfStatisticsCensusOfPopulationAndHousing2021FiveYear} \\
Gender & $b_i$ & 1: female, -1: male, uniform distribution & \\
Average Relationship duration for the node (in days) & ${\delta}_i$ & Poisson distribution with a mean of  $\langle\delta\rangle = 500$ &  \\
Lifetime Sexual Partners & ${lsp}_i$ & $T / {\delta}_i$ (rounded) & 
 \\
Cumulative sexual partners & ${sp}_i$ &  $[0, {lsp}_i)$ & 
\\
HPV-16 recovery period (in days) & $\alpha_i$ & Exponential distribution with a mean of 390 days & Olsen and Jepson \cite{Olsen2010HumanPapillomavirus} \\
HPV-16 transmissibility per coital act & $\beta$ & 0.3 & Olsen and Jepson \cite{Olsen2010HumanPapillomavirus} \\
Scale-free exponent & $\gamma$ & & \cite{Law2020PlacementMatters, Bell2017NetworkGrowth}\\
Scale-free fitness & $R^2$ & & \\
Node Degree & $k_i$ & & \\
Age difference between nodes & $\eta_{i,j}$ & $g_i - g_j$ & \\
Expected Relationship duration & $\Delta_{i,j}$ & & \\
Relationship age & $z_{i,j}$ & & \\

     & $ $ & &   \\

\textbf{Parameter} & $ $ & &   \\
     & $ $ & &   \\

Total timesteps & $T$ & 9000 & \\
Population  size & $N$ & 3000 & \\

Initial number of links  & $m_0$ & 5 s & \\

Frequency of intercourse & $f$ & $1/2$ during the first two weeks \newline $1/7$ after the first two weeks & Althaus et al. \cite{Althaus2012TransmissionChlamydia} \\

Scale-free calibration parameter & $\epsilon$ & 0.5 & This can be calibrated.  \\

Links added per joining node & $m$ &  & \cite{Law2020PlacementMatters, Bell2017NetworkGrowth}\\
Removed links per node time step & $m_r$ & & \\
Link removal rate per node per time step & $\theta$ & $1 / \langle\Delta\rangle$ & \\
Secondary links added per time step & $m_i$ &  &  (\ref{il1}), (\ref{il2})  
\\

Average degree & $\langle k \rangle$ & & \\
Assortativity & $r$ & & \cite{Law2020PlacementMatters, Lizier2011FunctionalStructural, Piraveenan2009LocalAssortativity, Piraveenan2007InformationCloningScaleFree} \\
Average age difference & $\langle \eta \rangle$ & 3.5 & Conroy-Beam and Buss \cite{Conroy-Beam2019WhyAge} \\

\hline
\end{tabular}
\label{parameter}
\end{center}
\end{table*}

The model has a range of other variables and parameters, which are summarised in Table \ref{parameter}. When the model is initialized, it is assumed that individuals under 20 are virgins, and have  had no sexual partners. It is assumed that  women prefer older men and vice-versa,  with an average age difference difference $\langle \eta \rangle=3.5$ \cite{Conroy-Beam2019WhyAge}. The sexual contact network growth model is implemented with three mechanisms: (1) new node introduction; (2) link removals; (3) secondary link formation. The first mechanism represents new people (nodes) entering (becoming part of) the contact network (as opposed to the relevant populations, which they are assumed to be already members of). Naturally, they need to create links with existing nodes in order to become part of the network, and the parameter $m$ represents how many links  they make, on average, when they first become part of the sexual contact network.  In this respect this mechanism is essentially similar to classical scale-free network growth models such as the Barabasi-Albert model \cite{Barabasi1999EmergenceScaling} or the Bianconi-Barabasi model \cite{Bianconi2001CompetitionMultiscaling}. The second mechanism represents discontinuation of relationships between people who are already in the sexual contact network, with the possibility that a person may `leave' the network temporarily if they have no links (no active relationships) at a particular time. The third mechanism represents the process whereby new links (relationships) are created between existing nodes (people) in the contact network. Herein lies the significance of this model: it does not merely creates a scale-free network, which several existing growth models can do,  but it creates a heterosexual contact network by mimicking the real-world process of relationship formation and maintenance, which is unique to sexual contact networks. The growth model also consists of two phases: in Phase 1, all three mechanisms work simultaneously (that is, people are continuing to join the contact network), while in Phase 2, only the last two mechanisms work (whereby it is assumed that all people in the considered population have joined the network, and new relationships are formed only between people(nodes)  which have had at least one relationship previously). The following subsection describe the three mechanisms in more detail:

\subsubsection{Mechanism I - New Node Introduction} Similar to the Barabási -Albert model and the Bianconi-Barabási model, in the beginning $t_0 = 0$, there are $m_0$ bipartite edges connecting $m_0$ pairs of  nodes in the network, representing $m_0$ monogamous heterosexual relationships.  At each time step $t$, a new node $i$ is introduced to the network and linked to $m$ preexisting nodes, until there is no new node available. The selection of $m$ nodes to be connected to node $i$ is made preferentially, and depends on both degree and fitness of the existing nodes.

The fitness of node $j$ is defined as:
\begin{equation} \label{fitness}
    \phi_j = \frac{max\{\langle \eta \rangle, abs(g_i - g_j)\} * {abs(b_i - b_j)}}{max\{1, abs({lsp}_i - {lsp}_j)\} }
\end{equation}
where ${lsp}_i = T / {\delta}_i$ denoting the estimated Lifetime Sexual Partners (LSP) for node $i$. The pairing depends on node $i$ and $j$'s age, gender and LSP. The fitness of a node which is not of opposite gender is therefore zero. The probability of node $i$ choosing node $j$ is defined as:
\begin{equation} \label{probability}
    p_j = \frac{(k_j + \epsilon)\phi_j}{\sum_{h\in N}(k_h + \epsilon)\phi_h}
\end{equation}
where $k$ is the current degree of node $j$, and the existence of $\epsilon$ ensures that nodes  which have currently no links (due to all previous relationships being discontinued, for example) are still able to preferentially selected based on their fitness. Therefore  $\epsilon$  needs to be a positive real number, and it can be assigned a value to calibrate the growth model. Links created through this mechanism are called primary links hereafter,  to distinguish them from links that are created from mechanism III.

\subsubsection{Mechanism II -  Link Removals}
Several  growth models introduced in Section \ref{background} only consider forming connections; however, in this context,  not all relationships can last permanently. Thus, besides relationship formation, we have added a new mechanism for relationship discontinuation.

Therefore when each link is formed, it is  assigned with an expected relationship duration, $\Delta_{ij}$.  The expected duration of the relationship (edge) depends on the average  relationship duration assigned to both partners (nodes), such that:

\begin{equation}
    \Delta_{ij} \sim Exp(min\{{\delta}_i, {\delta}_j\})
\end{equation}

When the age of the link reaches its duration, the link will be removed, demonstrating the termination of the relationship. Therefore, at each time step $t$, the rate of edges to be removed $\theta$ is  $1 /\langle \Delta \rangle $, and the number of removed edges per node per timestep is written as $m_r = \theta  / N $.

\subsubsection{Mechanism III  -  Secondary Link Formation}
The Barabasi-Albert model \cite{Barabasi1999EmergenceScaling} and Bianconi-Barabasi model \cite{Bianconi2001CompetitionMultiscaling} introduced earlier did not involve the formation of secondary links among existing nodes in the network, though the  Caldarelli model \cite{Caldarelli2002ScaleFreeNetworks} did.  In the present context of sexual contact network formation, it is clear that two individuals who are  both already members of the sexual contact network may in time form a relationship with them, and this does not have to be limited to the time when either of them joined the network.

Thus, we propose a third mechanism  to create secondary links between existing nodes. We postulate that the creation of such secondary links will be at  a rate which matches the deletion of primary links. Therefore, after mechanism I is completed, the number of links will be stable in the network. This mechanism uses fitness-based preferential attachment similar to Mechanism I  to create secondary links between nodes which are already part of the contact network. It is assumed, without loss of generality,  that secondary links are initiated by female nodes. The probability of a female node initiating a secondary link (note well that this should not be taken to mean `second' relationship -  the female node may have any number of primary links (relationships) already) is defined as

\begin{equation}
    q_i = \frac{max\{0, ({lsp}_i - {sp}_j)\}}{\sum_{h\in N}max\{0, ({lsp}_h - {sp}_h)\}}
\end{equation} which means that a female node (person) who has the greatest difference between their expected lifetime sexual partners (defined as the inverse of their average relationship period, $\delta_i$), and the cumulative sexual partners they have already had, is the most likely to initiate secondary relationships. The probability of  a male node being selected for this relationship follows a fitness-based preferential attachment mechanism, similar to that in mechanism I.

As mentioned before, the growth model has two phases.  In Phase 1, all three mechanisms operate simultaneously. Therefore, at each time step $t$, the number of secondary links $m_i$ is set as:

\begin{equation} \label{il1}
    m_i = \frac{(mt + m_0)\theta}{N}
\end{equation}

During phase 2, where only Mechanism II and Mechanism III operate, the number of secondary links per timestep is

\begin{equation} \label{il2}
    m_i = \frac{M\theta}{N}
\end{equation}

\begin{figure}[ht]
    \centerline{\includegraphics[scale = 0.9]{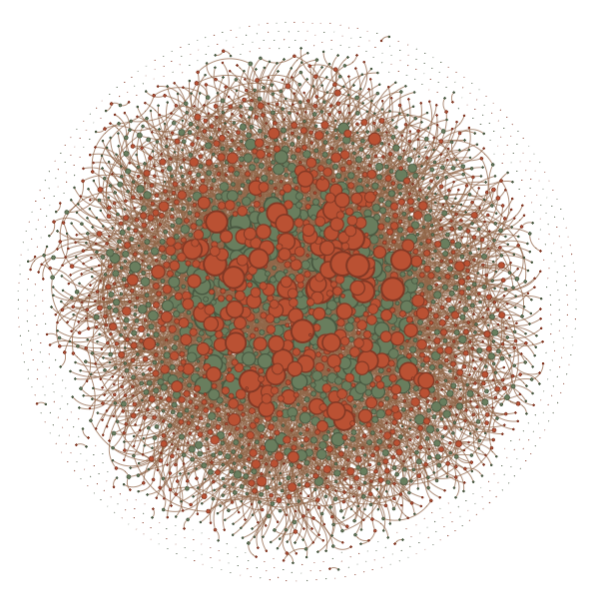}}
    \caption{A Heterosexual Contact network grown by the SeCoNet growth model, with  $N = 3000$ and average degree $\langle k \rangle = 3.28$. The average relationship duration for nodes is  $ \langle \delta \rangle = 500$. Node size is proportional to node degree. Red: Females, Green: Males.}
    \label{visual}
\end{figure}

Fig. \ref{visual} shows a heterosexual contact network grown by the SeCoNet growth model.

\subsection{The HPV Transmission Model} \label{hpv}

The HPV Disease dynamics transmission model is implemented as an SIR epidemic model. The population is compartmentalized into three compartments: (1) Susceptible; (2) Infectious; (3) Recovered, in terms of health status. There is no vaccination implemented in the model, and all recovered individuals are assumed to have lifetime immunity after recovery. Each individual of the population and their sexual contacts are directly represented in the network as nodes and links. The underlying topology represents the heterogeneity in sexual contacts and the transmission dynamics of the infection. 

The implemented infection  dynamics in the model is specific to HPV-16 variant, which is responsible for  approximately 50\% of cervical cancer diagnosis \cite{Lowy2008HumanPapillomavirus}. In addition to the parameters defined in the sexual contact network growth model,  we define epidemic parameters, including population level parameters such as frequency of sexual acts $f$, transmissibility of HPV-16 per sexual act $\beta$, and  individual recovery period $\alpha_i$. These are also included in Table \ref{parameter}.  Althaus et al. \cite{Althaus2012TransmissionChlamydia} suggests that in a relationship, the frequency of sexual intercourse decreases from once every two days to once a week after the first two weeks; therefore, the rate of sexual contact  $f$ each day is set as 0.5 in the first 14 time steps then becomes $1/7$ afterwards. Note that we assume the disease transmissibility as a global parameter,  while we assume that recovery period varies from person to person. As the sexual contact network grows, infection is seeded once for every 100 timesteps: that is, one person  out of every hundred people joining the contact network is assumed to be infectious.

The initial simulations run for 9000 days. The purpose of this simulation, at this stage, is to demonstrate the suitability of the SeCoNet growth model in simulating HPV disease dynamics..

\section{Results and Discussion} \label{results}

Here we present the simulation results of our model.  The simulation results consists of (1) topological analysis of the networks grown from the SeCoNet growth model (2) Disease dynamics analysis on the grown networks.  The population $N$ is set as 3000, the initial network consists of $m_0 = 5$ pairs of heterosexual nodes, and the average relationship duration is 500 days,  as shown in Table \ref{parameter}.  In the figures presented below, each data point represents the average of 10 simulation runs unless otherwise stated.

\subsection{Topological Analysis of grown networks}

\begin{figure}[ht]
    \centerline{\includegraphics[scale = 0.32]{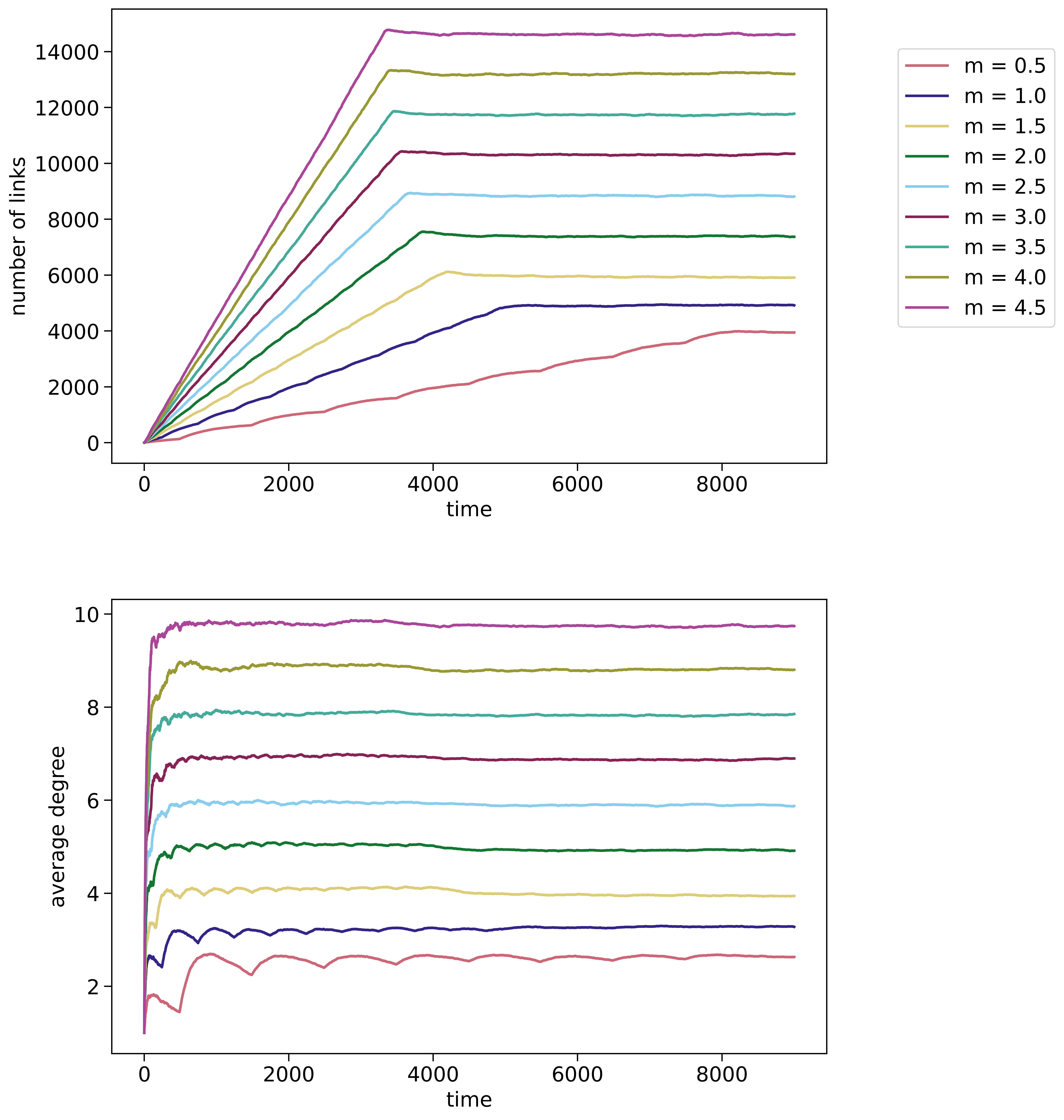}}
    \caption{Number of links and average degree against time  in SeCoNet growth model for different $m$ values. Here $N = 3000$, $m_0 = 5$ and $\langle \delta \rangle = 500$ }
    \label{size}
\end{figure}

 Fig \ref{size} shows the growth of the network (in terms of relationships) for different $m$ values, which, it can be recalled, is a growth parameter representing the number of primary links created when each node joins the network. A similar parameter is used in well-known scale-free network growth models such as the Barabasi-Albert model.  It can be seen that the network growth speed and the final size of the network is positively correlated with the value of $m$. Also phase 1  and phase 2 of the SeCoNet growth model are clearly visible in the figure, whereby in phase 1 the number of links increase rapidly, while in phase 2 the number of links (relationships) stabilise. Even though the figure represents $9000$ timesteps,  the growth can be further simulated to represent the time length under consideration, and it is clear that in phase 2, the topological parameters change little, even though the actual relationships continue to change. It is also clear that the average-degree of the network at steady state (in phase 2) is closely correlated to the parameter $m$, as can be expected, even though it is important to note that due to the interplay between the three mechanisms of the SeCoNet model, the parameter $m$ does not directly represent average degree of the resultant network. Further, it can be seen that  higher values  of $m$ accelerates the introduction of new nodes, which shortens the duration of Phase 1.

\begin{figure}[ht]
    \centerline{\includegraphics[scale = 0.35]{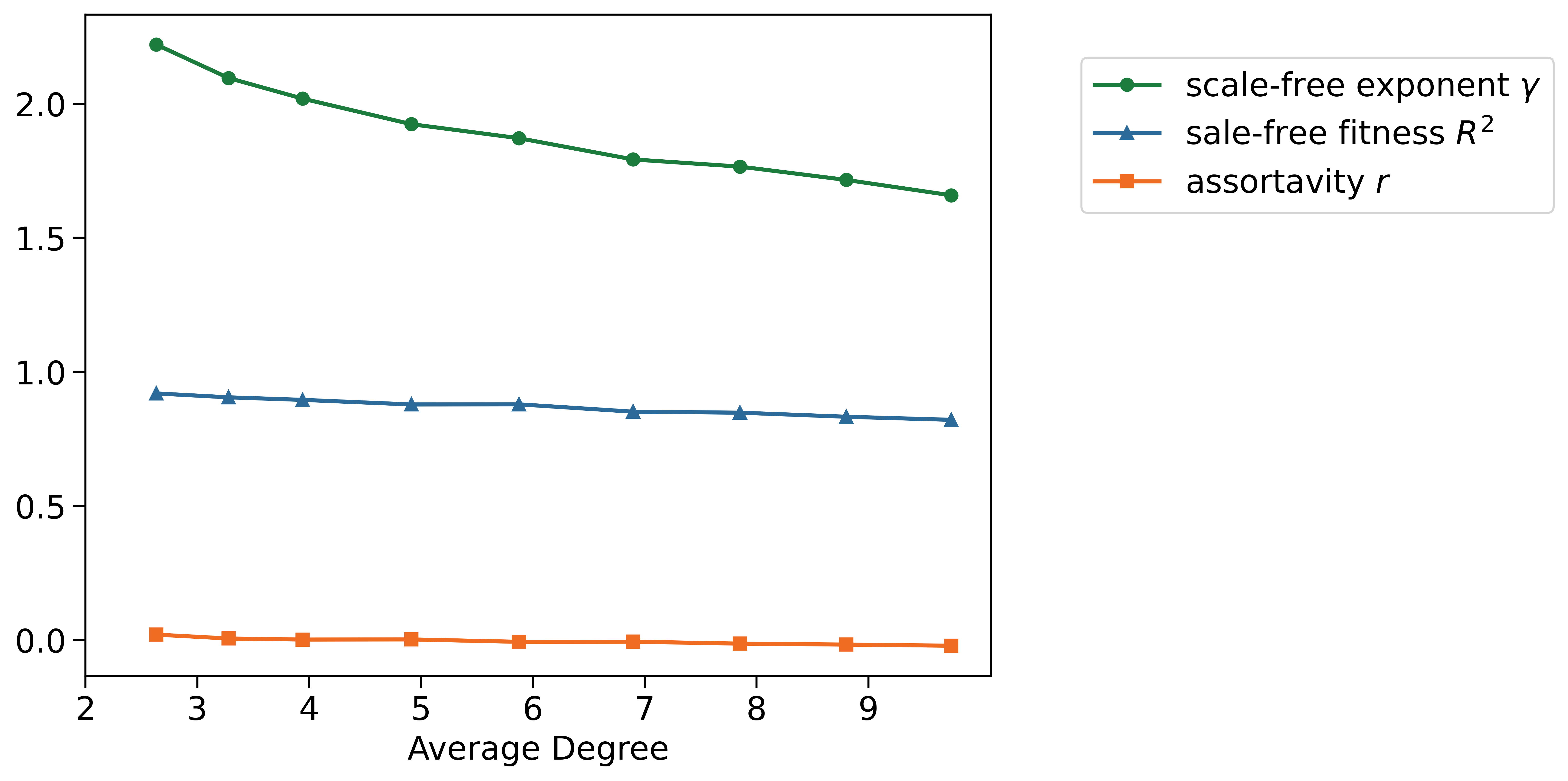}}
    \caption{Topological attributes (scale-free exponent, scale-free fitness, and assortativity) of networks grown by SeCoNet growth model. $N = 3000$, $m_0 = 5$ and $\langle \delta \rangle = 500$. }
    \label{sf}
\end{figure}

Fig. \ref{sf} shows the  scale-free exponent, scale-free fitness, and assortativity of a set of grown networks against average degree of these networks. The high scale-free fitness values (ranging from 0.95 to 0.9)  demonstrate that the generated networks exhibit scale-free topology. The scale-free exponent $\gamma$ varies between 2.5 and 1.5, which closely matches the scale-free exponents empirically observed in contact networks \cite{Kiss2006InfectiousDisease}.  Furthermore, as average degree increases,  the scale-free exponent $\gamma$ decreases. In particular,  when average degree is higher than $3$, the scale-free exponent is less than 2.0. For completeness we also show the degree assortativity of the networks, though it can be seen that the  degree assortativity is close to zero and does not vary with average degree. That is, the nodes do not show preferential mixing in terms of degrees.  This is not surprising since the intrinsic fitness of nodes would heavily influence preferential mixing, and even though nodes with higher degrees may be preferred by other nodes, those nodes have no incentive not to  develop relationships (make links) with less connected nodes: that is, degree-based preferential mixing occurs but is often one-sided, thus it does not influence degree-based assortativity. Nevertheless,  attribute-based assortativity (the so-called `scalar' assortativity \cite{Newman2003MixingPatterns}) could well be prevalent in the grown networks, and this is one aspect of the grown networks which needs further analysis.

\begin{figure*}[ht]
    \centering
    \includegraphics[scale = 0.28]{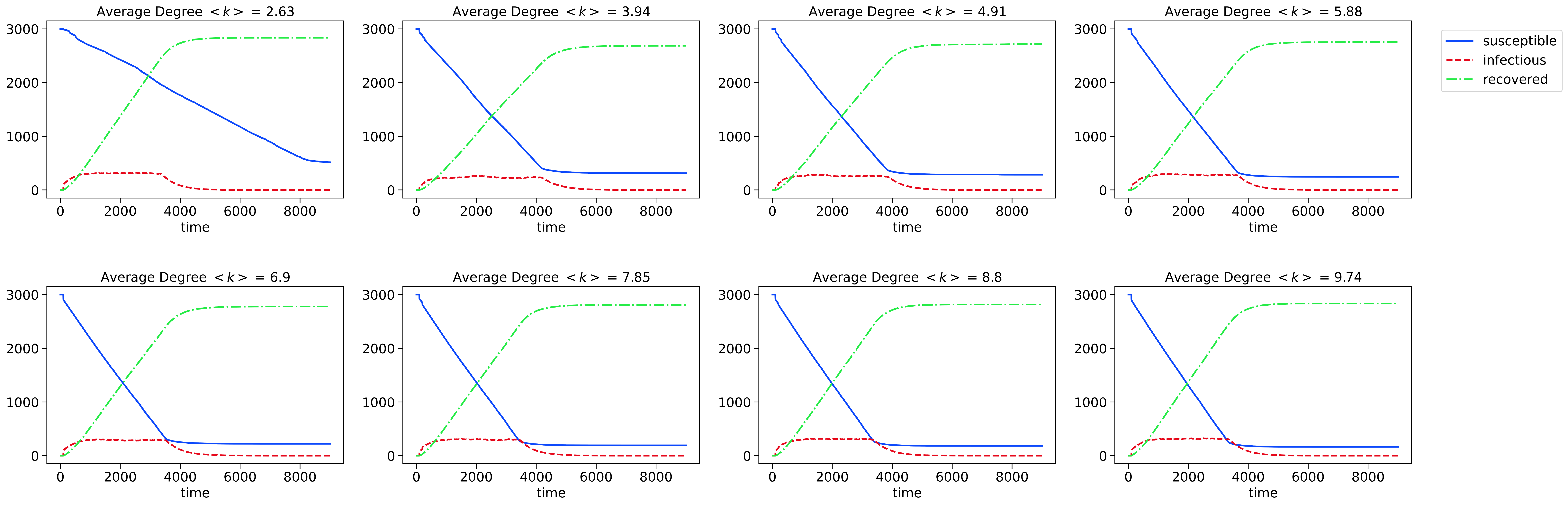}
    \caption{HPV-16 disease dynamics on networks grown by SeCoNet growth model. Here, transmission probability per sexual act $\beta = 0.3$, mean infection duration $\langle \alpha  \rangle= 390$ days. Infection is seeded by introducing one infectious node into the network every $1000$ timesteps while the network is growing in phase 1.}
    \label{sir}
\end{figure*}

\subsection{Simulating HPV Disease dynamics}

To confirm that the networks evolved from the SeCoNet growth model are suitable to analyse HPV disease dynamics, we conducted a set of basic infection transmission simulations on these contact networks.  The results of these experiments are shown in Fig. \ref{sir}, for a range of networks with differing average degrees. It can be seen that in all-cases, the classic dynamics of SIR models is observed, that is, the susceptible proportion reduces rapidly, the infectious proportion peaks then reduces, and then the recovered proportion increases rapidly,  even though the population is not-well mixed and contacts are represented and governed by the underlying contact network topology.  It can also be observed that the peak amplitude of the infectious proportion increases when average degree increases. That is, the higher connectivity seems to encourage a more vigorous infection transmission, as expected. It can also bee seen that the network saturates more quickly with recovered people when connectivity increases, due to the higher infection peaks. In all cases though, it could be noted that a very small proportion of the populace remains infected at the end of the simulation. This is due to the long average  recovery period of HPV, which results in people with longer recovery periods (outliers in the recovery period distribution) taking years to recover.  In short, all grown networks demonstrate expected disease dynamics.

It is important to stress here that the purpose of these simulations is not to discover something new about HPV disease dynamics. Rather, it is to demonstrate the utility of the SeCoNet growth model as a tool to capture the growth of HPV contact networks. Essentially, the main contribution of the present work is to introduce this growth model, while future work will focus on using this growth model to gain new insights about HPV disease dynamics, and treatment and vaccination strategies.

\section{Conclusion}
In this work, we introduce the SeCoNet (Sexual Contact Network) growth model, which produces scale-free contact network models using three mechanisms which reflect the real world relationship forming processes of people. The growth model is focused on heterosexual relationships and is calibrated to Australian demographics at present, though it can easily be adapted to other demographics. It is intended for simulation of Human Papillomavirus (HPV) infection. The growth happens in two phases, in the first of which all three growth mechanisms are at work, and the number of relationships rapidly increases, while in the second phase only two mechanisms are at work, and the number of relationships is fairly stable, though the topology continues to change and
evolve. Our analysis shows that the model creates highly scale-free networks, as intended, and growth parameters can be adjusted to produce scale-free networks with a desired scale-free exponent. We demonstrate that a compartmental HPV infection transmission model can be effectively simulated on networks produced by the SeCoNet growth model. 

It is important to note here that this is not another growth model to create scale-free networks in generic terms. Scale-free network growth models are many. The uniqueness of this model is that it uses realistic relationship building processes
in the specific context of sexual contact networks to generate realistic sexual contact network models, which happen to be scale-free as empirically observed. Therefore, the SeCoNet growth model should be primarily seen as a sexual contact network growth model, rather than a growth model to create scale-free networks in general. 

It should be noted however that even though the SeCoNet growth model is inspired by, and at present calibrated to, HPV infection, the most common Sexually Transmitted Infection (STI), the model could be easily adapted to simulate other STIs, such as Chlamydia, Gonorrhea, and Syphilis. It can also easily be extended to cover non-heterosexual contacts and relationships. Similarly, the SeCoNet growth model can be calibrated to be used in understanding HPV spread in any demographic or region. As mentioned earlier, high-risk HPV infection poses risks of complications like cervical cancer to significant sections of female populations in regions such as East Africa. Therefore, the SeCoNet growth model will be widely useful to computational epidemiologists. 

It should also be noted that in this work we primarily presented the growth mechanisms of the model, and undertook some basic analysis of HPV transmission on this model for the purposes of demonstrating the growth model’s utility. Future work involves using this model to extensively simulate HPV transmission and vaccination uptake, and their interdependency with the contact network topology. We will in particular look at vaccine uptake decision making of individuals and the emergence of herd immunity and how that depends on contact network topology. The growth model presented here is the necessary first step for such analysis, and particularly useful in contexts where empirical data about sexual contacts is rare. The model can similarly be adapted, with some calibration, to analyse vaccination uptake in the context of other STIs.

\bibliographystyle{IEEEtran}
\bibliography{main}

\end{document}